\documentclass[conference]{IEEEtran}

\ifCLASSINFOpdf
   \usepackage[pdftex]{graphicx}
   \graphicspath{{./png/}{./pdf/}{./jpeg/}}
   \DeclareGraphicsExtensions{.pdf,.jpeg,.png}
\else
   \usepackage[dvips]{graphicx}
   \graphicspath{{./eps/}}
   \DeclareGraphicsExtensions{.eps}
\fi
\usepackage{amssymb}
\usepackage{amsmath}
\usepackage{array}
\ifCLASSOPTIONcompsoc
  \usepackage[caption=false,font=normalsize,labelfont=sf,textfont=sf]{subfig}
\else
  \usepackage[caption=false,font=footnotesize]{subfig}
\fi
\usepackage{color,epsfig,calc}
\usepackage{balance}

\usepackage{algorithm}
\usepackage{algpseudocode}
\usepackage{acro}
\usepackage{amssymb}
\usepackage{amsmath}
\DeclareMathOperator*{\argmax}{arg\,max}
\DeclareMathOperator*{\argmin}{arg\,min}
\usepackage{tikz}
\usepackage{textcomp}
\usepackage{hyperref}
\usepackage{lipsum}
\newcommand\copyrighttext{%
  \footnotesize \textcopyright 2020 IEEE. Personal use of this material is permitted.
  Permission from IEEE must be obtained for all other uses, in any current or future 
  media, including reprinting/republishing this material for advertising or promotional 
  purposes, creating new collective works, for resale or redistribution to servers or 
  lists, or reuse of any copyrighted component of this work in other works. 
  }
\newcommand\copyrightnotice{%
\begin{tikzpicture}[remember picture,overlay]
\node[anchor=south,yshift=10pt] at (current page.south) {\fbox{\parbox{\dimexpr\textwidth-\fboxsep-\fboxrule\relax}{\copyrighttext}}};
\end{tikzpicture}%
}

\begin{document}

\title{A Blind Beam Tracking Scheme for\\ Millimeter Wave Systems}

\newcommand*{\affaddr}[1]{#1} 
\newcommand*{\affmark}[1][*]{\textsuperscript{#1}}
\newcommand*{\email}[1]{\texttt{#1}}

\author{Steve Blandino\affmark[1]$^{,}$\affmark[2],
Thibault Bertrand\affmark[3],
Claude Desset\affmark[2],
Andr\'{e}  Bourdoux\affmark[2],
Sofie Pollin\affmark[1]$^{,}$\affmark[2],
J\'{e}r\^{o}me Louveaux\affmark[3]\\
\affaddr{\affmark[1]Department of Electrical Engineering, \ KU Leuven, Belgium}\\
\affaddr{\affmark[2]imec, Belgium}\\
\affaddr{\affmark[3]Institute of Information and Communication Technologies, Electronics and Applied Mathematics,\\ UC Louvain, Belgium}\\
}

\maketitle
\copyrightnotice

\begin{abstract}
Millimeter-wave  is one of the technologies  powering the  new generation of wireless communication systems.  To compensate the high path-loss,
millimeter-wave devices need to use highly directional antennas. Consequently, beam misalignment causes strong performance degradation  reducing the link throughput
or even provoking a complete outage.
Conventional solutions, e.g. IEEE 802.11\,ad, propose the usage of additional training sequences  to track beam misalignment. These methods however introduce significant overhead  especially in dynamic scenarios.
In this paper we propose a beamforming scheme that can reduce this overhead. First, we propose an algorithm to design a codebook suitable for mobile scenarios. Secondly, we propose a blind beam tracking algorithm 
 based on particle filter, which describes the angular position of the devices
with a posterior density function constructed by  particles. The proposed scheme reduces by more than 80\% the overhead caused by  additional training sequences.
\end{abstract}


\IEEEpeerreviewmaketitle

\section{Introduction}
Millimeter-wave (mm-wave) wireless communication is considered the enabling technology
of next-generation  wireless systems. More than
20 GHz of spectrum is available at mm-wave
 to accommodate the ever-increasing throughput requests thanks also to the
advances on high-speed electronics enabling wireless systems operating at  high carrier frequency and with wide modulation bandwidth.

 The  propagation path loss at mm-wave requires the use of  large number of antennas to concentrate the radiated power into  narrow beams. Consequently,  the millimeter wave channel is  sparse in space  and beam misalignment causes strong signal to noise ratio (SNR) drops reducing the link throughput or even provoking  a complete outage. Efficient beamforming strategies, which can correctly  align the beams  even in dynamic  scenarios,  are of  crucial importance  for mm-wave devices to guarantee ubiquitous coverage and  the required  throughput.
 
IEEE 802.11\,ad/ay \cite{80211ad} provides explicit training sequences (TRN)  appended at the end of data packets
to support  a  beam tracking procedure (TRN-BT).
These training sequences can be  used to monitor the SNR and trigger a beam-sweep whenever
a strong drop in SNR occurs. During the  beam-sweep phase, both transmitter and receiver  sequentially send TRN sequences  on each antenna sector to find the steering providing the highest SNR. 
 These procedures introduce latency, which leads to throughput reductions. 
 Recently, algorithms  predicting  the device  motion have been proposed to  proactively adapt the beam pattern before an SNR drop occurs \cite{kang2018millimeter,Zhang2016}.
 
 Beyond the tracking schemes, the design of a robust code-book including wide beams,  less sensitive to the device motion, can significantly reduce the training overhead at the cost of lower SNR. 
 The analysis of a beamforming strategy, which include both code-book design and beam tracking for dynamic scenario is considered in \cite{7366948}, where a beamforming strategy was proposed by combining the codebook design with  an anticipatory movement prediction algorithm that utilizes mobile device’s sensors (accelerometer and magnetometer) to accurately forecast its next location.

In this paper, we present  a beamforming strategy leveraging on both a wide beam and a blind motion prediction, reducing the triggering of beam-sweep events.
The design of the proposed beamforming scheme provides  two main contributions:
\begin{itemize}
    \item Beam pattern design: we propose an algorithm to generate a code-book suitable for dynamic scenarios.  The key idea of the proposed beam pattern is to enlarge the beamwitdh while minimizing the secondary lobes. 
    \item Blind beam-tracking: we propose a predictive beam-tracking algorithm based on particle filtering, which enables an accurate beamforming  in dynamic scenarios, reducing by more than 80\% the triggering of   beam-sweep procedures. 
\end{itemize}

The reminder of the paper is organized as follows: Section \ref{sec:system_model} presents the system model including the signal model, the frame structure and the channel model. Section \ref{sec:codebook} describes how to properly design the beam pattern in both static and dynamic scenarios. Section \ref{sec:pf} introduces the proposed blind beam-tracking algorithm. Section \ref{sec:simulation}  presents the performance of the proposed beamforming scheme.  Finally, Section \ref{sec:conclusion} gives the concluding remarks.

\section{System Model}
\label{sec:system_model}
In this section, we present the system model used in the
paper. We describe  the  structure of the considered mm-wave communication system including signal model, frame structure and channel model.

\subsection{Signal Model}
Consider a mm-wave system comprising a fixed access point (AP) with $M$ antennas and one moving user station (STA)  with $N$ antennas.
Assume that the  antennas, in both devices,  are organized as uniform linear array (ULA) composed of isotropic radiators each provided with a phase shifter.
While in this  paper  we focus on the design of the beam for a single 
user system,  these schemes are valid for the analog beamforming design of  hybrid MIMO architectures  usually  used in  multi-user scenarios \cite{8472217}.

Assume a wideband  single carrier system (SC-FDE) as in 802.11\,ad/ay.  The
circulant property satisfied by using the cyclic prefix permits to describe  the  downlink transmission  as follow:
\begin{equation}
    y[k] = \mathbf{f}_\mathrm{STA}^\mathrm{H}\mathbf{H}[k] \mathbf{f}_\mathrm{AP} x[k] + n,
\end{equation}
where $x[k]$ is the equivalent transmitted symbol at the sub-carrier $k$,  $\mathbf{f}_\mathrm{AP}$ and $\mathbf{f}_\mathrm{STA}$  are  the $\mathbb{C}^{M\times 1}$ and the $\mathbb{C}^{N\times 1}$ the antenna weight vectors (AWV), i.e. frequency flat analog beamformer and combiner vectors of the AP and STA respectively.  $\mathbf{H}$  is the $\mathbb{C}^{N\times M}$ frequency selective  channel.
$n$ is the additive Gaussian noise  with variance $\sigma_n$.

\subsection{Frame structure}
\begin{figure}
  \centering
   \medskip
  \includegraphics[width=3.5in]{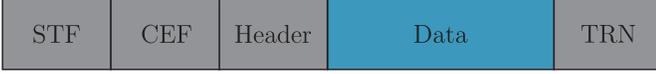}
   \caption{802.11\,ad frame structure. TRN field is an optional overhead to enable beam-tracking.}
  \label{fig:frame}
\end{figure}

AP and the STA  communicate using  two different frame structures:
\begin{enumerate}
    \item Beam training frames contain pilot sequences, e.g. Golay sequences, which are used in the first phase of the communication  when the communication link is not established or when an  outage occurs and the link needs to be restored. The initial access procedure is successful when $\mathbf{f}_\mathrm{AP}$ and $\mathbf{f}_\mathrm{STA}$ are acquired.
    \item Data frames, in Figure \ref{fig:frame}, are used when the link is established. At  the beginning of the data frame, shorter pilot sequences compared to the beam training frames (STF, CEF) are present to provide synchronization, SNR and  channel estimation or  CFO and phase noise correction. Optional training  overhead (TRN) can be appended at the end of  a packet to allow beam training procedures. 
\end{enumerate}
 A finite set of vectors is usually available  since the
transceivers need to quickly switch between the different vectors, hence predefined
vectors are generally stored in a codebook to reduce computation complexity and  power consumption.
In the following   we propose a robust code-book  for dynamic scenarios from which  $\mathbf{f}_\mathrm{AP}$ and $\mathbf{f}_\mathrm{STA}$ are selected. Devices motion however makes the initial $\mathbf{f}_\mathrm{AP}$ and $\mathbf{f}_\mathrm{STA}$ outdated.  Our beam-tracking  algorithm uses the mandatory overhead (STF and CEF), without adding the additional overhead at the end of the packet (TRN) to update $\mathbf{f}_\mathrm{AP}$ and $\mathbf{f}_\mathrm{STA}$.

\subsection{Channel Model}

Given the   sparse nature of the mm-wave channel in time and space a quasi-deterministic channel model \cite{Weiler2016} is used to represent realistic multi-path components. 
The design of the beam-tracking schemes presented in this paper relies on  the line-of sight (LOS)  component of the channel, 
since at mm-wave only the LOS component  can provide reliably
 high transmission rate.
The description of the normalized $N \times M$ LOS channel is given by:
\begin{equation}
    \bar{\mathbf{H}} = \sqrt{MN}\, \alpha \mathbf{a}_r(\phi_r) \mathbf{a}_t(\phi_t)^H,
\end{equation}
where $\alpha$ is the complex gain of the channel at the distance $d$ and  $\phi_t$ and   $\phi_r$ are the azimuth angles of departure and arrival respectively, which depend on the relative position of the devices.
The vectors $\mathbf{a}_r$ and $\mathbf{a}_t$ are the normalized receive and transmit array  steering  vectors respectively, which  incorporate all of the spatial characteristics of the array.
For an $N$-elements uniform linear array oriented along  the $y$-axis, the array steering  vector $\mathbf{a}$ is:
\begin{equation}
\mathbf{a} = [ 1\,     e^{j2\pi\frac{d}{\lambda}\sin{\phi}}\,     e^{j2\pi\frac{2d}{\lambda}\sin{\phi}}\,  \cdots\,   e^{j2\pi\frac{(N-1)d}{\lambda}\sin{\phi}}]^T,
\end{equation}
where $\lambda$ is the carrier wavelength $d$ is the inter-element spacing. We omit the description of the elevation since an ULA has omni-directional pattern in the elevation dimension.
\section{Beam Codebook Design}
\label{sec:codebook}
Large multi-antenna transceivers focus their radiations to a narrow region, hence  the beam should  be accurately tuned to exploit the  beamforming gain. In this section we describe how to  properly design a code-book    to maximize the SNR in both static and dynamic scenarios.

\subsection{Beam Codebook for Static Devices}
\label{subsec:static}
In the case of static devices  the design of  narrow pencil beams  optimizing the SNR is desired.
This is possible by adapting the phase coefficients
 $\mathbf{f}_\mathrm{STA}$ and $\mathbf{f}_\mathrm{AP}$  such that:
\begin{equation}
\label{optim:static}
 \begin{aligned}
 (\mathbf{f}_\mathrm{STA}^\mathrm{opt}, \mathbf{f}_\mathrm{AP}^\mathrm{opt}) = &  \underset{\mathbf{f}_\mathrm{AP},\mathbf{f}_\mathrm{STA}}\argmax 
& & |\mathbf{f}^H_\mathrm{STA}\bar{\mathbf{H}}\mathbf{f}_\mathrm{AP}|^2  ,\\
& \text{subject to}
& & \mathbf{f}_\mathrm{AP} \in \mathcal{F}_\mathrm{AP},\quad ||\mathbf{f}_\mathrm{AP}|| = 1\\
& & & \mathbf{f}_\mathrm{STA} \in \mathcal{F}_\mathrm{STA},\quad ||\mathbf{f}_\mathrm{STA}|| = 1
\end{aligned}
\end{equation}
where $\mathcal{F}_\mathrm{AP}$ and  $\mathcal{F}_\mathrm{STA}$ are the feasible set  of the AP  and STA beamsteering vectors.

With full channel knowledge  a simple solution is based on
singular value decomposition. Letting the channel's singular value
decomposition be $\bar{\mathbf{H}} = \mathbf{U}\mathbf{\Sigma}\mathbf{V}$, the optimal vectors are $\mathbf{f}_\mathrm{AP}^\mathrm{opt} = \mathbf{U}^{(1)}$ and $\mathbf{f}_\mathrm{STA}^\mathrm{opt} = \mathbf{V}^{(1)}$, the right and left eigenvector associated with the strongest eigenvalue.
However since the sparse nature of the mm-wave channel, the
optimal SVD beamforming vectors are  given  by  the array steering vector in the strongest direction \cite{6292865}  that is:
\begin{equation}
    \begin{aligned}
        \mathbf{f}_\mathrm{STA}&= \mathbf{a}_t(\phi_t),\\
        \mathbf{f}_\mathrm{AP}&=\mathbf{a}_r(\phi_r)^H,
    \end{aligned}
\end{equation}
hence the estimation of $\phi_t$ and $\phi_r$ are sufficient to solve the problem in (\ref{optim:static}).

\subsection{Beam Codebook for Dynamic Devices}
For a large number of antennas, in the case of a scenario with moving STAs the 
approach presented in \ref{subsec:static} may not  be desirable  since a  small mismatch between the estimation and the true value of the angle directions  can cause a drastic SNR drop.

To guarantee  a more reliable link even with a large number of antennas, we modify  problem (\ref{optim:static}) such that the SNR is not maximized only for the combination ($\phi_t$,$\phi_r$) but it is maximized for the set of angles  ($\Phi_t$,$\Phi_r$), in which 
$\Phi_t~=~[\phi_t -\Delta\phi_\mathrm{bw}/2, \phi_t +\Delta\phi_\mathrm{bw}/2 ]$ and 
$\Phi_r~=~[\phi_r -\Delta\phi_\mathrm{bw}/2, \phi_r +\Delta\phi_\mathrm{bw}/2 ]$, with  
$\Delta\phi_\mathrm{bw}$ indicating the beam-width.
In this case the optimization problem is given by:
\begin{equation}
\label{eq:optim_cb}
 \begin{aligned}
 (\mathbf{f}_\mathrm{STA}^\mathrm{opt}, \mathbf{f}_\mathrm{AP}^\mathrm{opt}) = &  \underset{\mathbf{f}_\mathrm{AP},\mathbf{f}_\mathrm{STA}}\argmax 
& & |\mathbf{f}^H_\mathrm{STA}\bar{\mathbf{H}}\mathbf{f}_\mathrm{AP}|^2  ,\\
& \text{subject to}
& & \mathbf{f}_\mathrm{AP} \in \mathcal{F}_\mathrm{AP},\quad ||\mathbf{f}_\mathrm{AP}|| = 1\\
& & & \mathbf{f}_\mathrm{STA} \in \mathcal{F}_\mathrm{STA},\quad ||\mathbf{f}_\mathrm{STA}|| = 1\\
& & & \Delta\phi_\mathrm{bw}=C.
\end{aligned}
\end{equation}
where $C$ is a constant.

The problem is solved separately for the AP and the STA, e.g. the codebook of the AP is designed assuming an omni-directional receiver STA. A solution of the problem is provided by using  a conventional evolutionary algorithm.
 The main idea is to translate problem (\ref{eq:optim_cb}) into a fitness function. The AWV with higher fitness in
the population will have higher chances of survival. At each iteration a new generation of AWV is introduced by  modifying the AWV of the previous generation.
 After few generations, the selected AWV is  likely to have high fitness values and therefore represent
a (sub-)optimal solution.
To solve this problem, we minimize  the following fitness function $\mathbb{F}_\mathrm{AP}$:
\begin{equation}\label{eq:fitfct}
 \begin{aligned}
    \mathbb{F}_\mathrm{AP}(   {\mathbf{f}_\mathrm{AP}}) &= \underbrace{\frac{1}{\Delta\phi_\mathrm{bw}}\int_{\phi \in \Phi_t} (A_{m} - |A(\phi)|)^{2}d\phi}_{\mathbb{F}_{1}(   {\mathbf{f}_\mathrm{AP}})} +\\ &\beta_1\underbrace{\frac{1}{\pi-\Delta\phi_\mathrm{bw}}\int_{\phi \in \Bar{\Phi}} |A(\phi)|^{2} d\phi}_{\mathbb{F}_{2}(   {\mathbf{f}_\mathrm{AP}})} - \\
    &\beta_2 \underbrace{\frac{1}{\Delta\phi_\mathrm{bw}}\int_{\phi \in \Phi_t} (|A(\phi)|)^{2}d\phi}_{\mathbb{F}_{3}(   {\mathbf{f}_\mathrm{AP}})} 
    \end{aligned}
\end{equation}
where  $\Bar{\Phi} = [-\pi/2    , \phi_t-\Delta\phi_\mathrm{bw}/2] \cup [ \phi_t+\Delta\phi_\mathrm{bw}/2, \pi/2    ]$ is the region outside the beamwidth, $A(\phi)  = \mathbf{a}_t \mathbf{f}_\mathrm{AP}^H $, $A_m = \mathbb{E}\left\{|A(\phi)\right|\}$ averaged in $\Phi_t$. $\beta_1$ and $\beta_2$ are  arbitrary real scalars used for weighting $\mathbb{F}_{1}(   {\mathbf{f}_\mathrm{AP}})$, $\mathbb{F}_{2}(   {\mathbf{f}_\mathrm{AP}})$ and $\mathbb{F}_{3}(   {\mathbf{f}_\mathrm{AP}})$.
The fitness function in equation (\ref{eq:fitfct}) refers to the codebook design of the  AP, however a similar function can be written for the STA.
The fitness function has been chosen to minimize the array gain in $\bar{\Phi}$ (= $\mathbb{F}_{2}(   {\mathbf{f}_\mathrm{AP}})$), minimize the level of ripples of $|A(\phi)|$ in $\Phi$ (= $\mathbb{F}_{1}(   {\mathbf{f}_\mathrm{AP}})$) and maximize the array gain in  $\Phi$(= $\mathbb{F}_{3}(   {\mathbf{f}_\mathrm{AP}})$).

Algorithm \ref{algo:BeamWidening} can be summarized as follow. In the first stage 
a random population of AWV $\mathbf{F}_\mathrm{AP}\in \mathbb{C}^{N\times N_P}$, where $N_P$ is the size of the population, is generated. The coefficients  of $\mathbf{F}_\mathrm{AP}$ have unitary amplitude but a random phase. At each iteration, the best individual (in the sense of the minimisation of $\mathbb{F}(   {\mathbf{f}_\mathrm{AP}})$) is selected and a new population is created from the perturbation of this individual.

  \begin{algorithm}
   \caption{Beam Codebook design}\label{algo:BeamWidening}
   \begin{algorithmic}[1]
     \Require {$N, N_{P}, N_\mathrm{max}, \eta_\mathrm{max}$}
     \State Initialize population $\mathbf{F}_\mathrm{AP}$
     \State Initialize perturbation amplitude $\eta =1$
     \State Initial value of fitness function min\_value = $+\infty$
       \While{$\eta < \eta_\mathrm{max} $}
              \While{$n < N_\mathrm{max} $}
        \State{find $\mathbf{f}_\mathrm{STA}^\mathrm{opt}$} that solves: 
        \begin{equation*}
        \begin{aligned}
        \mathbf{f}_\mathrm{AP}^\mathrm{opt} = &  \underset{\mathbf{f}_\mathrm{AP}\in \mathbf{F}_\mathrm{AP}}\argmin 
            & & \mathbb{F}(\mathbf{f}_\mathrm{AP}) ,\\
        \end{aligned}
        \end{equation*}
        \If{min\_value $ > \mathbb{F}(\mathbf{f}_\mathrm{AP}^\mathrm{opt})$}
        \State min\_value $\gets \mathbb{F}(\mathbf{f}_\mathrm{AP}^\mathrm{opt})$
        \State $n\gets N_\mathrm{max}$
        \Else
        \State $n\gets n+1$
        \If{$n== N_\mathrm{max}$ }
        \State $\eta \gets 2\eta$
        \EndIf
        \EndIf
        \State Perturb phases of $\mathbf{f}_\mathrm{AP}^\mathrm{opt}$ with random coefficient to create new population   $\bar{\mathbf{F}}_\mathrm{AP}$
         \State $\mathbf{F}_\mathrm{AP} = [ \mathbf{f}_\mathrm{AP}^\mathrm{opt}, \bar{\mathbf{F}}_\mathrm{AP}]$ 
       \EndWhile
              \EndWhile
       \State \textbf{return} $\mathbf{f}_\mathrm{AP}^\mathrm{opt} /||\mathbf{f}_\mathrm{AP}^\mathrm{opt} ||$
   \end{algorithmic}
 \end{algorithm}
 
 \begin{figure}
  \centering
   \medskip
  \includegraphics[width=3in]{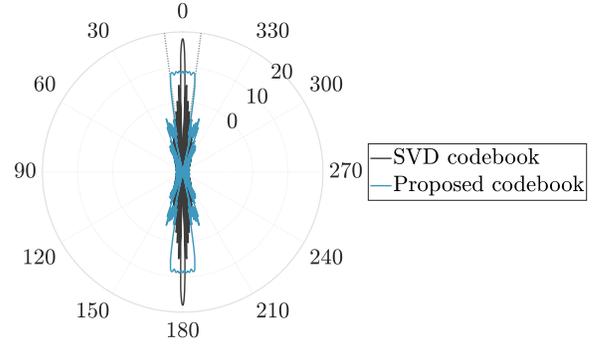}
   \caption{Array pattern for $\phi_t = 0$ for both SVD and proposed codebooks assuming 64 antennas. The proposed codebook enlarge the beamwidth such that $\Delta\phi_\mathrm{bw}=15^\circ$. Array gain is expressed in
dB.}
  \label{fig:beam15}
\end{figure}
 
Figure \ref{fig:beam15} shows an example of beam pattern when the AP uses 64 antennas for $\phi_t =0$ generated by designing the codebook according with algorithm 1. A larger beam-width is obtained compared to a conventional SVD codebook to prevent SNR drops in dynamic scenarios. 

\section{Beam Tracking Schemes}
\label{sec:pf}
The SNR of the  link usually decreases over
time due to the devices mobility. When the SNR becomes too low, generally below a threshold, it becomes
necessary to find a more suitable AWV. Instead of performing an exhaustive beam search using additional TRN sequences, predictive beam-tracking schemes that can anticipate the device motion,  can be used. In this section we first review the conventional beam-tracking  used in 802.11\,ad systems. Then, we propose a novel beam tracking scheme based on particle filtering.

\subsection{Training Sequence Beam Tracking (TRN-BT)}

A procedure named beam tracking has been included in the
IEEE\,802.11ad  specification and allows a fast beam refinement.
 This beam tracking procedure is a request/response
procedure. Transmitter training (TRN-T) and receiver training (TRN-R) fields are appended
at the end of data packets so that the STAs can train their transmit and receive AWVs.
During the transmission of a TRN-R field, the same transmit AWV must be used
for the transmission of all TRN subfields so that the receiver can train its receive
AWV. Inversely, during the reception of a TRN-T field, the same receive AWV must
be used so that the transmitter can train its transmit AWV.

\subsection{Predict and track with Particle Filter (PF-BT)}


To estimate the user motion, and predict the best AWV without sending TRN sequnces, we propose a tracking scheme based on particle filter (PF). The  PF can estimate the past, current or future states of a Markov process using  noisy and partial observations. 

Consider the Markov process:
\begin{equation}\label{eq:Markov_proc_true}
    \begin{split}
       	\theta_{t} &= \theta_{t-1} + \dot{\theta}_{t-1}\Delta{t}\\
	\dot{\theta}_{t} &= f(\dot{\theta}_{t-1}; \nu_{t})\\
        \gamma_{t} &= g(\theta_{t};w_{t})\\
    \end{split}
\end{equation}
where $\theta_{t}$ is the angular position and $\dot{\theta}_{t}$ is the angular velocity of the user every $\Delta{t}$\,ms. $\theta_{t}$ and $\dot{\theta}_{t}$ represent the state $\mathbf{\Theta}_t$ of the Markov process, while $\gamma_t$ is the observed  SNR.
 $f$ and $g$ are functions describing the process evolution. $\nu_ {t}$ and $w_{t}$ are random variables.
 We would like to estimate recursively the posterior distribution of the hidden state $p(\mathbf{\Theta}_{t}|\gamma_{1:t-1})$ using a PF. The problem can be addressed recursively in two steps \cite{Intro_MonteCarlo}:
\begin{itemize}
    \item Prediction: 
    \begin{equation}
    \begin{split}
    p(\mathbf{\Theta}_{t}|\gamma_{1:t-1}) &= \int p(\mathbf{\Theta}_{t},\mathbf{\Theta}_{t-1} |\gamma_{1:t-1})d\mathbf{\Theta}_{t-1}  \\ &= \int p(\mathbf{\Theta}_{t}|\mathbf{\Theta}_{t-1})p(\mathbf{\Theta}_{t-1}|\gamma_{1:t-1})d\mathbf{\Theta}_{t-1}.
    \end{split}
    \end{equation}
    \item Update: 
        \begin{equation}
        \begin{split}
    p(\mathbf{\Theta}_{t}|\gamma_{1:t}) &= \dfrac{p(\gamma_{t}|\mathbf{\Theta}_{t})p(\mathbf{\Theta}_{t})|\gamma_{1:t-1})}{p(\gamma_{t}|\gamma_{1:t-1})} \\ &= \dfrac{p(\gamma_{t}|\mathbf{\Theta}_{t})p(\mathbf{\Theta}_{t}|\gamma_{1:t-1})}{\int p(\gamma_{t}|\mathbf{\Theta}_{t})p(\mathbf{\Theta}_{t}|\gamma_{1:t-1})d\mathbf{\Theta}_{t}}.
    \end{split}
    \end{equation}
\end{itemize}
The PF represents the probabilistic distribution of the hidden state by particles. Each particle can be seen as one possible evolution of the hidden state and the density of the particles describes the probabilistic distribution of the hidden state. A given hidden state is more likely if many particles have a state  close to it. 
 \begin{figure}
     \centering
     \includegraphics[width=3.5in]{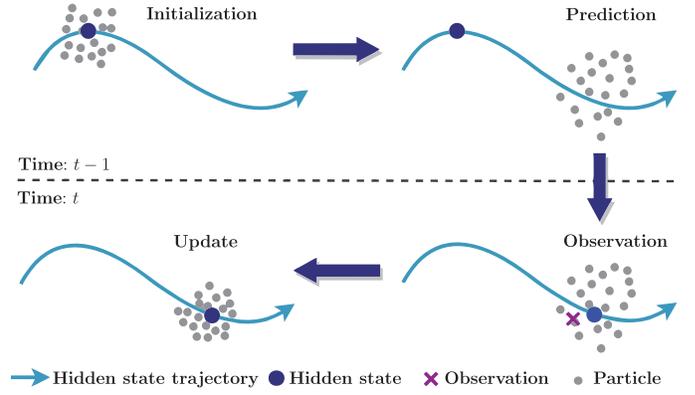}
     \caption{Illustration of the particle filter.}
     \label{fig:Filt_principle}
 \end{figure}
The particle filter operations are illustrated in Figure \ref{fig:Filt_principle}. At each iterations, the prediction and update steps are performed.
The prediction step uses the probabilistic distribution of the hidden state $p(\mathbf{\Theta}_{t-1}|\gamma_{1:t-1})$ and based on the conditional probabilistic state evolution $p(\mathbf{\Theta}_{t}|\mathbf{\Theta}_{t-1})$ estimates the most probable state at the next time instant
applying the function $f_{t}(\mathbf{\Theta}_{t-1}, \nu_{t})$ to each particle. The update step estimates the most probable state given all the observations up to time $t$ by using the predicted distribution $p(\mathbf{\Theta}_{t}|\gamma_{1:t-1})$ and the conditional probabilistic state observation $p(\gamma_{t}|\mathbf{\Theta}_{t})$.
Particles are selected according to $p(\gamma_{t}|\mathbf{\Theta}_{t})$ and are then resampled. A resampling operation  ensures that a sufficient number of particles can describe the probabilistic distribution of the hidden state.

The evolution of the angular speed $\dot{\theta}_{t}$ of the user needs to  consider two different scenarios:
\begin{itemize}
	\item The user motion at time $t$ follows the one at time $t-1$:  $f(\dot{\theta}_{t-1}; \nu_{t}) =  \dot{\theta}_{t-1} + \nu_{t}^{1}$. In this case, $\nu_{t}^{1}$ can be considered as a random Gaussian variable expressing the angular acceleration of the user.
	\item The user motion completely change between time $t-1$ and time $t$:  $f(\dot{\theta}_{t-1}; \nu_{t}) = \nu_{t}^{2}$. In this case, $\nu_{t}^{2}$ can be considered as a random variable taken from a uniform distribution.
\end{itemize}
The two  scenarios can have different  probability of occurrence, hence during the prediction step, two unequal sets of particles can  be considered such that each set is updated  with  a different distribution.

\section{Simulation}
\label{sec:simulation}
\begin{figure}
  \centering
   \medskip
  \includegraphics[width=2in]{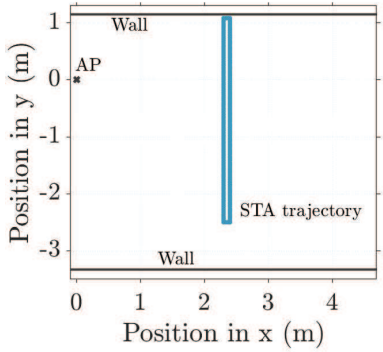}
   \caption{Indoor scenario with a fixed AP and a moving STA.}
  \label{fig:scenario}
\end{figure}
Let us consider an indoor scenario where a single moving STA is
communicating with a stationary AP. We assume that the antenna array of the STA is
made of a single antenna element, without losses of generalities, and we test our proposed schemes at the AP. The AP is provided with  an antenna array of $M= 64$ antennas. Let us also consider that the STA can move  with a
speed of 5\,km/h~=~1.39\,m/s and follows the trajectory depicted in Figure \ref{fig:scenario}.
It  is assumed that the AP knows perfectly the initial
angular position of the STA  and thus the beam is correctly steered  toward the STA. 

\begin{figure}
  \centering
   \medskip
  \includegraphics[width=3.45in]{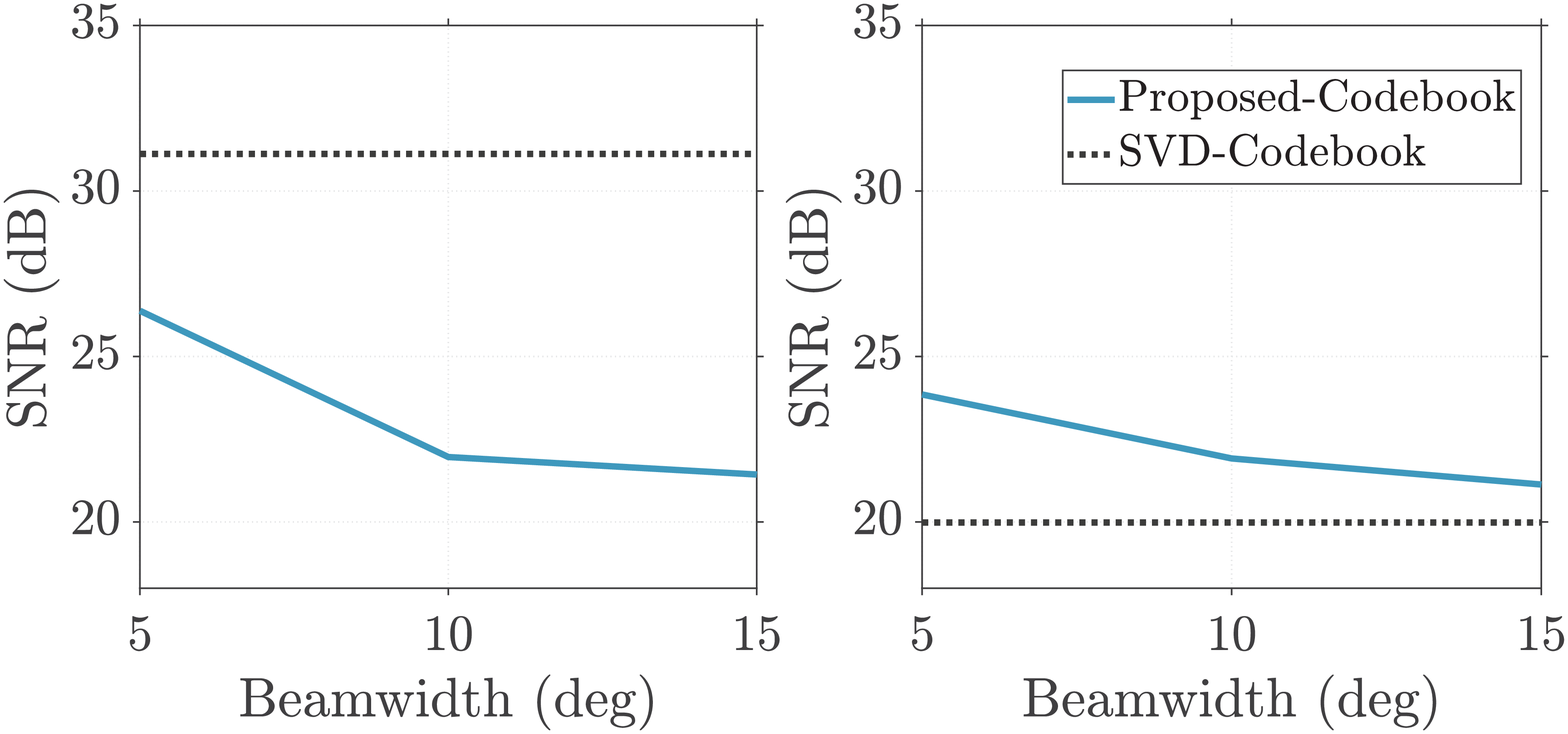}
  \\[-3ex]\subfloat[\label{fig:static}  Static]{ \hspace{.45\linewidth}}
          \subfloat[\label{fig:ch_6_ee_b32} Dynamic]{\hspace{.5\linewidth}}
   \caption{SNR vs beamwidth in static and dynamic scenarios.}
  \label{fig:snrbw}
\end{figure}
First, we test our codebook design in both static and dynamic scenarios.
We show the results for $\beta~=~2,\, N_P~=~100,\, N_\mathrm{max}~=~200 \text{ and } \eta_\mathrm{max}~=~10^5$. 
Figure \ref{fig:snrbw} shows the simulated average SNR, when the  STA follow the trajectory  making  2 complete loops. The SNR is presented   varying $\Delta\phi_\mathrm{bw}$, the beamwidth of our designed codebook according to Algorithm 1.  The figure shows also the SNR achieved assuming a conventional SVD based codebook.   In dynamic scenarios the design of a wider beam allows to achieve higher average SNR with an improvement of 4\,dB compared to a traditional SVD based coodebook when $\Delta\phi_\mathrm{bw}=5^\circ$. As expected instead in  a static scenario the SVD codebook achieves the  optimal performance. 
These results suggest that future mm-wave devices need to use codebooks, which include both very directional beams to maximize the SNR in static scenario, but also wider beams are needed to guarantee high throughput even in mobile scenarios.

\begin{figure}
  \centering
   \medskip
  \includegraphics[width=2in]{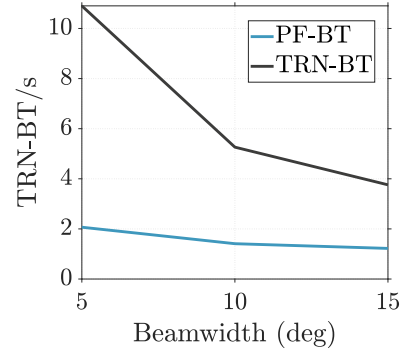}
   \caption{Frequency of TRN-BT procedure against  beamwidth with and without PF-BT.}
  \label{fig:tracking}
\end{figure}
Second, we analyze the performance of the proposed predictive beam tracking.
We assume that  sudden motion changes are less likely than a continuous movement, hence we assume that
98\% of the  particles update their angular velocity accordingly with   $\nu_{t}^{1}$ while the rest with  $\nu_{t}^{2}$. The choice of the  distribution $\nu_{t}^{1}$ and $\nu_{t}^{2}$ can be related with  the maximal angular speed of the user.  We assume the angular speed of the user  to be smaller than $\dot{\theta}_{max}$.
Considering a  user  moving with an absolute speed inferior to 1.5m/s, standing constantly at more than 1\,m from the AP, the maximal angular speed of the user is $\dot{\theta}_{max} =\mathrm{atan}\left(\frac{1.5}{1}\right) \approx 56^\circ /s.$
$\nu_{t}^{1}$ can be described  as a random variable taken from a zero mean Gaussian distribution of standard deviation $\dot{\theta}_{max}\Delta{t}$  meaning that about  68\% of the particles will accelerate by less than $56^\circ /s^2$ between $t-1$ and $t$.  $\nu_{t}^{2}$ can be described  as a random variable taken from a uniform distribution in the range $[-\dot{\theta}_{max} ; \dot{\theta}_{max} ]$.
When an SNR drop is experienced a TRN-BT is performed. The AP can   transmit TRN training sequences  every 20\,ms to test different AWVs. 
Figure \ref{fig:tracking} shows the number of TRN-BT/s, that is the frequency of triggering a beam sweep procedure by varying the codebook beam-width. When the AP performs particle filtering  for tracking the received SNR and predicting the user motion, the frequency of a TRN-BT procedure is reduced by more than 80\%. 
By increasing  the codebook beam-width the
number of TRN-BT/s reduces since  the user  stays longer inside the beam. 

\begin{figure}
  \centering
   \medskip
  \includegraphics[width=3.45in]{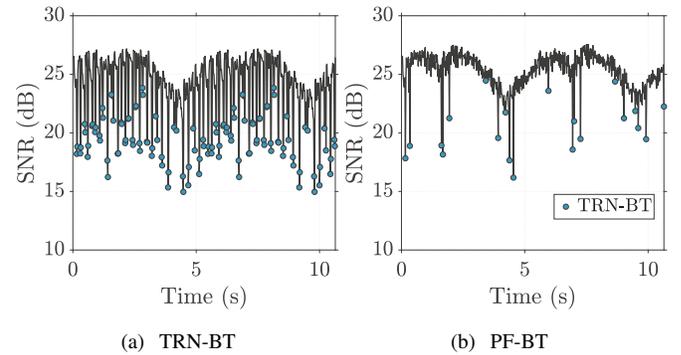}
  \\[-3ex]\subfloat[\label{fig:snr_time_a}  TRN-BT]{ \hspace{.45\linewidth}}
          \subfloat[\label{fig:snr_time_b} PF-BT]{\hspace{.5\linewidth}}
   \caption{SNR temporal evolution with and without PF-BT}
  \label{fig:snr_time}
\end{figure}
Finally, Figure \ref{fig:snr_time} shows the temporal evolution of the SNR for a user in motion. 
When beam-tracking relies only on TRN, the SNR drops continuously. When the AP predict the motion of the STA using PF-BT, the SNR experienced by the STA is much more stable with few interruptions.

\section{Conclusion}
\label{sec:conclusion}

This paper proposes a beamforming procedure, which leverages on two main contributions.
First, we propose an algorithm to design a beam codebook that offers a larger coverage compared to conventional solutions. The proposed codebook is  particularly relevant in dynamic scenarios in which an improvement of 4\,dB in the average SNR is achieved.  By increasing the codebook beam-width the user stays longer inside the main lobe experiencing less detrimental SNR drops.

Second, we propose a beam tracking algorithm based on particle filtering.  PF effectively predicts the user movement only with SNR measurements. Hence, PF can reduce more than 80\% the triggering of overhead pilots to re-align the beams.

As a future work,  a study considering different user motions
should be done to highlight the  robustness of the proposed algorithms. Moreover, experimental
validations  would be interesting to test the robustness of the proposed scheme with hardware non-idealities and a real propagation channel. Moreover, this scheme reduce the peak gain, hence another interesting question is the  maximum achievable range and the   highest modulation order achievable by the system.

\balance


\ifCLASSOPTIONcaptionsoff
  \newpage
\fi

\bibliographystyle{IEEEtran}
\bibliography{tracking_bib}

\end{document}